\documentclass[journal,twoside,web]{ieeecolor}
\usepackage{lcsys}
\usepackage{cite}
\usepackage{textcomp}

\usepackage{graphicx,algorithmicx,algorithm,algpseudocode,amsthm,amsmath,amsfonts,verbatim,mathtools,enumerate,bm,hyperref,dsfont,tikz,pgfplots,amssymb,cleveref,caption,subcaption,nicefrac}
\usepackage[acronym]{glossaries}
\usetikzlibrary{positioning}
\allowdisplaybreaks

\newcommand{\mnorm}[1]{{\left\vert\kern-0.25ex\left\vert\kern-0.25ex\left\vert #1 
    \right\vert\kern-0.25ex\right\vert\kern-0.25ex\right\vert}}

\newtheorem{definition}{Definition}

\newtheorem{proposition}{Proposition}

\newcommand{\ie}{{\it i.e.}}

\newcommand{\actprof}{\bm{a}}
\newcommand{\actprofset}{\mathcal{A}}

\newcommand{\corrSet}{\mathcal{G}}
\newcommand{\nashSet}{\mathcal{D}}
\newcommand{\nashConv}{\mathcal{H}}
\newcommand{\bigO}{\mathcal{O}}
\newcommand{\jointz}{\bm{z}}
\newcommand{\Occupy}{\mathrm{Occupy}}
\newcommand{\Yield}{\mathrm{Yield}}
\newcommand{\IntRange}[2]{\mathbb{N}_{[#1,#2]}}
\newcommand{\costFunc}{J}
\newcommand{\convHullMult}{\bm{\gamma}}
\newcommand{\lagMult}{\lambda}
\newcommand{\poisRate}{\upsilon}
\newcommand{\lagrangian}{\mathcal{L}}
\crefformat{equation}{(#2#1#3)}

\newacronym{kkt}{KKT}{Karush-Kuhn-Tucker}

\def\BibTeX{{\rm B\kern-.05em{\sc i\kern-.025em b}\kern-.08em
    T\kern-.1667em\lower.7ex\hbox{E}\kern-.125emX}}
\begin{document}
\title{Coordination in Noncooperative Multiplayer Matrix Games via Reduced Rank Correlated Equilibria}
\author{Jaehan Im, \IEEEmembership{Student Member, IEEE}, Yue Yu, \IEEEmembership{Member, IEEE}, David Fridovich-Keil, \IEEEmembership{Member, IEEE}, and Ufuk Topcu, \IEEEmembership{Fellow, IEEE}
\thanks{The authors would like to thank Dr. Sarah H.Q. Li for some helpful early discussions, and the anonymous associate editor and reviewers for their constructive feedback.}
\thanks{This work was supported by a National Science Foundation CAREER award under Grant No. 2336840 and a National Aeronautics and Space Administration ULI award under Grant No. 80NSSC21M0071}
\thanks{J. Im, and D. Fridovich-Keil are with the Department of Aerospace Engineering and Engineering Mechanics, The University of Texas at Austin, TX, 78712, USA (emails: jaehan.im@utexas.edu,\, dfk@utexas.edu). Y. Yu, and U. Topcu are with the Oden Institute for Computational Engineering and Sciences, The University of Texas at Austin, TX, 78712, USA (emails:  yueyu@utexas.edu,\, utopcu@utexas.edu).}}

\maketitle
\thispagestyle{empty}

\begin{abstract}

Coordination in multiplayer games enables players to avoid the lose-lose outcome that often arises at Nash equilibria. However, designing a coordination mechanism typically requires the consideration of the joint actions of all players, which becomes intractable in large-scale games. We develop a novel coordination mechanism, termed \emph{reduced rank correlated equilibria}. The idea is to approximate the set of all joint actions with the actions used in a set of pre-computed Nash equilibria via a convex hull operation. In a game with $n$ players and each player having $m$ actions, the proposed mechanism reduces the number of joint actions considered from $\bigO(m^n)$ to $\bigO(mn)$ and thereby mitigates computational complexity.
We demonstrate the application of the proposed mechanism to an air traffic queue management problem. Compared with the correlated equilibrium---a popular benchmark coordination mechanism---the proposed approach is capable of solving a problem involving four thousand times more joint actions while yielding similar or better performance in terms of a fairness indicator and showing a maximum optimality gap of 0.066\% in terms of the average delay cost. In the meantime, it yields a solution that shows up to 99.5\% improvement in a fairness indicator and up to 50.4\% reduction in average delay cost compared to the Nash solution, which does not involve coordination.

\end{abstract}


\section{Introduction}
\label{sec:introduction}
\IEEEPARstart{I}{n} a multiplayer game, each player minimizes its own cost by choosing its strategy---a probability distribution over its available actions---based on the other players' strategies. At a \emph{Nash equilibrium}, no player can reduce its cost by unilaterally changing its current strategy. Multiplayer games enable the modeling of noncooperative interactions in various applications, including air traffic management systems.

Coordination mechanisms in games allow players to avoid \emph{lose-lose outcomes}, which often occur in competitive interactions. The prisoner's dilemma and the chicken game \cite{Computing_correlated_equilibria_in_multi-player_games} are typical examples of such games. Coordination mechanisms help players avoid lose-lose outcomes by recommending joint actions to each player through a coordinator.

A \emph{correlated equilibrium} in a multiplayer game is a probability distribution over all joint actions of all players \cite{AUMANN_Corr} which provides a coordination mechanism as follows. Suppose a coordinator samples a joint action according to a correlated equilibrium known to all players and recommends it to all players. Then, each player cannot reduce its own cost by unilaterally deviating from the recommendation. Several studies have shown that the correlated equilibrium is useful for coordinating chicken games \cite{Computing_correlated_equilibria_in_multi-player_games}, 2-by-2 games \cite{coordinateCorr3}, Battle-of-Sexes games \cite{coordinateCorr1}, and public goods game \cite{coordinateCorr2}.

However, computing a correlated equilibrium becomes intractable as the number of players and actions increases, because it requires considering all joint actions. In a game with $n$ players where each player has $m$ actions, the number of all joint actions is $m^n$. This makes the computation of correlated equilibria intractable as $n$ and $m$ increase, which limits its usage in large-scale coordination problems \cite{Computing_correlated_equilibria_in_multi-player_games}.

\begin{figure}
    \centering
    \includegraphics[width=0.5\textwidth]{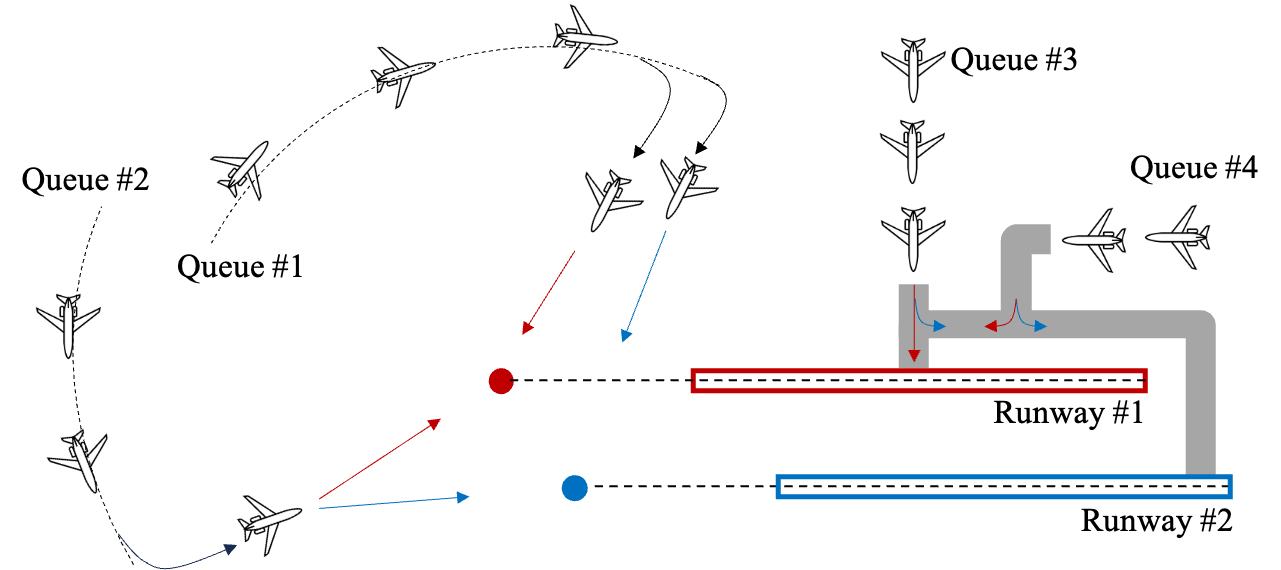}
    \caption{Illustration of a multiple departure-arrival queue coordination scenario. Each player (queue) may occupy both runways, a single runway, or yield to others.}
    \label{fig:introimage}
\end{figure}

We develop an algorithm that efficiently computes correlated equilibria based on a novel concept termed \emph{reduced rank correlated equilibria (RRCE)}. The set of RRCE inhabits the convex hull of multiple Nash equilibria, and this set approximates the set of all possible correlated equilibria. The proposed algorithm first computes multiple Nash equilibria and then obtains a set of RRCE. Unlike the correlated equilibrium, the computation of the Nash equilibrium only requires the consideration of the actions of individual players in isolation rather than all joint actions. As a result, the proposed algorithm reduces the number of joint actions considered during the correlated equilibrium computation from $\bigO(m^n)$ to $\bigO(mn)$. 

We evaluate the proposed algorithm by conducting a series of Monte Carlo simulations on an air traffic management problem, which typically requires coordination for efficient operation. The proposed algorithm solves a problem involving a four thousand times larger number of joint actions than a direct computation of the correlated equilibrium can solve. The proposed algorithm yields a solution that shows a $0.066\%$ gap (at worst) to the correlated equilibrium regarding the average cost per player. Yet our approach demonstrates superior fairness and average cost per player compared to a Nash solution that does not involve coordination. 

\section{Related Works} \label{section:RelatedWorks}
The challenge of obtaining the correlated equilibrium in large-scale problems is widely acknowledged \cite{Computing_correlated_equilibria_in_multi-player_games, ad_hoc_networks, SimpleApproxiamteEquilibria}. One research direction that addresses this issue is to utilize an evolutionary algorithm to obtain the correlated equilibrium \cite{Hart_adaptiveProcedureLeadingToCorr, EvoModelCorr, ad_hoc_networks}. These algorithms let players gradually converge to a correlated equilibrium through repeated games while players are set to follow a regret-minimization rule. However, this approach converges to a random correlated equilibrium, preventing the user from obtaining an equilibrium that fits their interests.

Learning dynamics is another approach to bypass the direct computation of the correlated equilibrium. This approach learns a strategy converging to a \emph{coarse correlated equilibrium}, a relaxed notion of correlated equilibrium, that minimizes the sum of players' costs \cite{EvoCorr, learningCCE, LearningEfficientCE}. However, a coarse correlated equilibrium is an equilibrium that does not guarantee that the players comply with the recommended actions, making it inappropriate to be applied for coordination purposes.

Various multiplayer games have extensively implemented a Nash equilibrium as coordination \cite{GameATM, NashBased_Highway, nashBasedControl1} and outcome prediction \cite{nashBasedControl2} tool. This approach computes an existing Nash equilibrium and uses it as a control command relayed to the players. If multiple Nash equilibria exist, the coordinator chooses the one that best fits the coordinator's interests \cite{GameATM}. However, since the Nash equilibrium does not consider joint actions, it generally yields a less efficient solution than the correlated equilibrium \cite{EvoCorr, LearningEfficientCE}.

Approximation of the correlated equilibrium has been attempted in mean-field games \cite{CorrApprox2_rev} and graphical games \cite{CorrApprox1_perf} to improve solution quality and reduce computation burden. However, to our knowledge, no research has focused on both (i) addressing the computational intractability of the correlated equilibrium and (ii) yielding a solution that fits the coordinator's interests.

\section{Nash and Correlated Equilibria 
\\ in Matrix Games}

\subsection{Matrix games}

We consider a game with \(n\in\mathbb{N}\) players in which each player has \(m\in\mathbb{N}\) actions. In this paper, we denote a ranged set of natural numbers $\{1,2,\ldots,n\}$ as $\mathbb{N}_{[1,n]}$. Let \(-i\) denote the set excluding the $i$-th element itself, \ie, \(\{{1,\ldots,n}\}\setminus i\). Lastly, let $v\in\mathbb{R}^n$ be a vector and let $[v]_k$ indicate the $k$-th element of $v$ for $k\in\mathbb{N}_{[1,n]}$.

\subsubsection{Player's action}
Let \(a_i\in\IntRange{1}{m}\) denote an action of player $i$ and \(\actprof:=\{a_1,\ldots,a_n\}\) denote a joint action of all players. Let $\actprofset$ be a set of all possible joint actions, $\actprof$. The cardinality of this set, $|\actprofset|$, is $m^n$.

\subsubsection{Player's strategy}
Let \(\Delta_m\coloneqq \{v\in\mathbb{R}^m_{\geq0}\mid v^\top \mathbf{1}_m=1\}\) and \(x_i\in\Delta_m\) denote player \(i\)'s \emph{strategy} as follows:
\begin{equation}
    \textstyle [x_i]_{a_i}\coloneqq \mathds{P}[\text{Player $i$ taking action $a_i$}].
\end{equation}
\subsubsection{Joint action distribution}
Let $\jointz\in\mathbb{R}^{m\times \overset{n}{\ldots} \times m}_{\geq 0}$ denote the \emph{joint action probability distribution} and \(\jointz_{\actprof}\in[0, 1]\) denote the probability of specific joint action profile \(\actprof\) occuring, \ie, 
\begin{equation}
    \textstyle\jointz_{\actprof} \coloneqq \mathds{P} [\text{Players take joint action }\actprof]=\mathds{P}[\actprof],
\end{equation}
where \( \sum_{\actprof} \jointz_{\actprof} = 1 \).

The joint action probability distribution profile $\jointz$ can be expressed in terms of the players' strategy $x_i$, \ie\ the conversion equation is shown below.
\begin{equation}\label{eq:actprof2jointactiondistOrig}
    \textstyle\jointz_{\actprof} = \prod_{i}[x_i]_{[\actprof]_i} .
\end{equation}
Note that this conversion is only valid from $x$ to $\jointz$. The reverse (decomposition of $\jointz$ to $x$) may not exist or result in indefinite strategy profiles $x$.

\subsubsection{Cost}
The cost that player \(i\) incurs by taking action \(a_i\) when player \(j\) takes action \(a_j\) (for all \(j\ne i\)) is given by \(\sum_{j\ne i}[C^{ij}]_{a_ia_j}\), where \([C^{ij}]_{a_ia_j}\) is the \(a_ia_j\)-th element (\(a_i\)-th row and \(a_j\)-th column) in the cost matrix \(C^{ij}\in\mathbb{R}^{m\times m}\).

The cost that player $i$ experiences is expressed as follows: 

\begin{equation} \label{eq:cost_with_Z}
    \textstyle c_i(\jointz) \coloneq \sum_{\actprof} \jointz_{\actprof} \sum_{j\in{-i}}[C^{ij}]_{[\actprof]_i [\actprof]_j}.
\end{equation}
Note that one can compute $c_i$ with individual player's strategy $x_i$ by transforming $x_i$ to $\jointz$ via \Cref{eq:actprof2jointactiondistOrig}.

\subsection{Nash equilibrium (NE)}
The Nash equilibrium is a set of individual strategies in which no player can reduce its cost by unilaterally changing its strategy. It exists regardless of the presence of coordination. 
\begin{definition}[Nash equilibrium] \label{def:NashEquilibrium}
    A Nash equilibrium is a strategy \(x^\star\) such that for all \(i \in \mathbb{N}_{[1,n]}\),
    \begin{equation} \label{eq:def:NashEquilibrium}
        \textstyle c_i(x_i^\star,x_{-i}^\star) \leq c_i(x_i,x_{-i}^\star),\,\forall x_i\in\Delta_m,\,x_i^\star\neq x_i.
    \end{equation}
\end{definition}
Under an appropriate constraint qualification, the \acrlong{kkt} (\acrshort{kkt}) conditions for all players must be satisfied at a Nash equilibrium \cite[p.26]{Pang}. The KKT conditions for player $i$ are
\begin{equation} \label{eq:NashKKT}
    \begin{split}
        \textstyle \nabla_{x_i}\lagrangian_i &= \textstyle \sum_{j\neq-i} C^{ij} x_j - \lagMult^i - \mu^i\mathbf{1}_m=\mathbf{0}_m, \\
        \textstyle \mathbf{1}_m^\top x_i&=1, \enskip \lagMult^{i\top}x_i=0, \enskip x_i \geq \mathbf{0}_{m}, \enskip  \lagMult^i \geq \mathbf{0}_m,
    \end{split}
\end{equation}
 where $\lagMult^i=[\lagMult_1^i,\ldots,\lagMult_m^i]^\top\in\mathbb{R}^m$ and $\mu^i\in\mathbb{R}$ are Lagrange multipliers, and $\lagrangian_i$ is player $i$'s Lagrangian.

\subsection{Correlated equilibrium (CE)}
A correlated equilibrium is a joint action probability distribution, $\jointz$, such that after a particular joint action profile $\actprof^\star$ is drawn from $\jointz$ and each action $a_i^\star\in\actprof^\star$ is recommended to each player $i$, playing $a_i^\star$ is the optimal choice for each player $i$. Here, we assume that the players know the distribution $\jointz$.

\begin{definition}[Correlated equilibrium] \label{def:CorrEquilibrium}
    A correlated equilibrium is a probability distribution $\jointz$ over the set of joint action profiles $\mathcal{A}$ such that
    \begin{equation} \label{eq:correlatedDef}
        \textstyle \mathbb{E}[c_i(a_i^\star)|a_i^\star,\jointz] \leq \mathbb{E}[c_i(a_i)|a_i^\star,\jointz],\,a_i^\star = [\actprof^\star]_i,
    \end{equation}
    where $a_i^\star \neq a_i$, for all $i\in\mathbb{N}_{[1,n]}$ and for all joint action profiles $\actprof\in\actprofset$.
\end{definition}

The expected cost that player $i$ experiences when the player chooses action $a_i$ with action suggestion $a_i^\star$ given $\jointz$, $\mathbb{E}[c_i(a_i^\star)|a_i^\star,\jointz]$, is denoted with $\Phi_{a_i^\star}^{\jointz}(a_i)$:
\begin{equation} \label{eq:corrRationalityCond}
    \textstyle \Phi_{a_i^\star}^{\jointz}(a_i) = \sum_{\hat{\actprof}} \jointz_{\hat{\actprof}}\sum_{j\ne{i}}[C^{ij}]_{[\actprof]_i[\actprof]_j},
\end{equation}
where $\hat{\actprof}\in\{\actprof\in\actprofset|[\actprof]_i=a_i^\star\}$. Then, the correlated equilibrium is a joint action distribution $\jointz$ that satisfies the following conditions:
\begin{equation} \label{eq:corrFormulation}
    \begin{aligned}
            & \textstyle \Phi_{a_i^\star}^{\jointz}(a_i^\star) \leq \Phi_{a_i^\star}^{\jointz}(a_i), \,\, \forall (a_i, a_i^\star) \in \IntRange{1}{m},\,\forall i \in \mathbb{N}_{[1,n]},\\
            & \textstyle \sum_{\actprof\in\actprofset} \jointz_{\actprof} = 1,\enskip \jointz_{\actprof} \geq 0, \,\, \forall \actprof\in\actprofset.
    \end{aligned}
\end{equation}

\section{Reduced Rank Correlated Equilibria Algorithm}

The correlated equilibrium computation becomes quickly intractable as the number of joint actions grows exponentially in the number of players ($n$). Recall that the number of joint actions considered to compute the correlated equilibrium is $\bigO(m^n)$ while for the Nash equilibrium, the figure is $\bigO(mn)$. This affects the computational complexity of solving for each equilibrium. Although computational complexity may vary according to which solver one uses, we provide an example when using the interior point method (IP).

\begin{proposition}[Number of linear equations to find NE and CE with IP]
    The computation of the Nash equilibrium and the correlated equilibrium with IP involves solving $n(m+1)$ and $2m^n+3m^2n+1$ linear equations, respectively.
\end{proposition}
\begin{prooff}
    The interior point (IP) method solves a system of linear equations at every iteration \cite[p.393]{Nocedal}.
    The Nash equilibrium is formulated as a linear complementarity problem with $n(m+1)$ linear equations when solved with IP \cite[p.1037]{Pang}.
    The correlated equilibrium is formulated as a linear program with $2m^n+3m^2n+1$ linear equations when solved with IP \cite[p.394]{Nocedal}.
\end{prooff}

To exploit this apparent difference in scaling, we propose an algorithm that approximates the set of correlated equilibria with multiple Nash equilibria.

\subsection{Correlated equilibrium set approximation}

Let $\corrSet:=\{\jointz|\jointz \text{ satisfies \Cref{eq:corrFormulation}}\}$ denote a set of correlated equilibria. Since \Cref{eq:corrFormulation} comprises only linear constraints, $\corrSet$ is a convex polytope. In addition, the set of all Nash equilibria is a subset of $\corrSet$ \cite{CENEGemetry}. Therefore, the convex hull of Nash equilibria is a subset of the correlated equilibria.

Let a set with $d$ distinct Nash equilibria denoted as $\mathcal{D}$, with the superscript $k$ indicating the $k$-th Nash equilibrium
\begin{equation}
    \textstyle [\nashSet]_k\coloneq\{x_1^k,x_2^k,\ldots,x_n^k\}, k\in\IntRange{1}{d}.
\end{equation}
It is possible to convert $[\nashSet]_k$ into a joint action probability distribution $\jointz$ using \Cref{eq:actprof2jointactiondist}. Let $\jointz^k\in\mathbb{R}^{m\times \overset{n}{\ldots} \times m}_{\geq 0}$ denote the joint action probability distribution of the $k$-th equilibrium in $\nashSet$ such that
\begin{equation}\label{eq:actprof2jointactiondist}
    \textstyle
    \jointz_{\actprof}^k = \prod_{i}[x_i^k]_{[\actprof]_i}
\end{equation}
for all \( \actprof\in\actprofset\). The convex hull of $\jointz^{k\in\IntRange{1}{d}}$ is a subset of the correlated equilibrium set $\corrSet$ \cite{CENEGemetry, ConvPol}, \ie\
\begin{equation} \label{eq:convHull}
    \textstyle
    \nashConv \coloneq \mathrm{conv}(\jointz^k) \subseteq \corrSet,
\end{equation}
where the convex hull of $\{\jointz^k\}_{k=1}^d$ is denoted as $\nashConv$ and $\mathrm{conv}(\jointz^k)=\{\sum_{i=1}^d [\convHullMult]_i\jointz^i\mid\convHullMult\in\mathbb{R}_{\geq 0}^d,\,\mathbf{1}^\top_d\convHullMult = 1\}$.

\begin{definition}[Rank of a tensor] \label{def:tensorRank}
    A simple tensor $T^s$ is a tensor that can be expressed with the outer products of $p$ vectors $v$ where $p$ is the dimension of a tensor, \ie\, $[T^s]_{i,\overset{p}{\ldots},k}=[v_1]_i\ldots [v_p]_k$. $T$ is of rank $t$ if and only if $t$ is the smallest number of simple tensors required to express $T$ \cite[p.309]{bourbaki}, \ie\, $T=T_1^s+\ldots+T_t^s$.
\end{definition}

The joint action probability distribution of the Nash equilibrium, $\jointz_{\actprof}^k$, has a rank of 1 as the Nash equilibrium is expressed with an outer product of the player's strategies $x_i$ as shown in \Cref{eq:actprof2jointactiondist}, and is thus a simple tensor. Since all Nash equilibria are elements of the set of correlated equilibria $\corrSet$ \cite{CENEGemetry}, an element in $\corrSet$ with equal or higher rank than any elements in $\nashConv$ always exists.
Based on this characteristic, elements within the set $\nashConv$ are termed \emph{reduced rank correlated equilibria (RRCE)}.

\subsection{Multiplayer coordination and RRCE algorithm}
We consider an optimization problem that seeks a correlated equilibrium $\jointz$, which minimizes a particular cost function $\costFunc$ that considers the social context. By recommending actions sampled from the correlated equilibrium to each player, a coordinator can help the players reach a joint action that minimizes $\costFunc$. Such a socially optimal joint action is often unachievable without a coordination mechanism.

The cost function $\costFunc$ that evaluates the quality of a correlated equilibrium---such as efficiency and fairness---is denoted as
\begin{equation}
    \textstyle \costFunc:\mathbb{R}^{m\times \overset{n}{\ldots} \times m}_{\geq 0}\to\mathbb{R}.
\end{equation}
In particular, we search for an \emph{optimal} correlated equilibrium by solving the following optimization:
\begin{equation} \label{eq:CorrOptimization}
    \begin{array}{ll}
         \textstyle\underset{\jointz}{\mbox{minimize}} &\costFunc(\jointz) \\
         \textstyle \mbox{subject to}  & \text{condition from \Cref{eq:corrFormulation}}. \\
    \end{array}
\end{equation}
We denote this method as the \emph{Correlated Equilibrium (CE) algorithm}. As stated previously, solving this problem requires high computation costs. 

In contrast, we propose an algorithm that computes RRCE, termed the \emph{RRCE algorithm}. The RRCE algorithm consists of two phases. First, the algorithm searches for multiple Nash equilibria---each satisfies \Cref{def:NashEquilibrium}---and computes the corresponding joint action distributions according to \Cref{eq:actprof2jointactiondist}. The computed joint action distribution is denoted \(\{\jointz^1, \ldots, \jointz^d\}\). We suggest two methods for searching Nash equilibria: (i) \emph{random initialization} method, which randomly initializes the numerical solver \cite{juliaPATH} when solving \Cref{eq:NashKKT}, and (ii) \emph{brute-force} method, which exhaustively enumerates all deterministic joint actions and identifies all pure Nash equilibria \cite{BruteSurvey}.

Second, the RRCE algorithm computes a reduced rank correlated equilibrium given by \(\sum_{k=1}^d [\convHullMult^\star]_i \jointz^k\), where \(\convHullMult^\star\in\mathbb{R}^d_{\geq 0}\) is the optimal solution of the following optimization problem: 
\begin{equation} \label{eq:RRCEOptimization}
\begin{array}{ll}
     \textstyle \underset{\convHullMult}{\mbox{minimize}} &\costFunc(\sum_{i=1}^d [\convHullMult]_i\jointz^i) \\
     \textstyle \mbox{subject to}  & \convHullMult\geq\mathbf{0}_d,\enskip \mathbf{1}^\top_d\convHullMult = 1.
\end{array}
\end{equation}

\section{Coordination in Air Traffic Management}
A noncooperative multiplayer coordination problem is frequently observed in air traffic management. Consider a situation where two aircraft are trying to land at an airport with a single runway. In this case, the players are aircraft, and the limited resource is runway occupancy. Each aircraft prefers to occupy the runway (\ie\, land) immediately rather than wait until the other aircraft lands and vacates the runway. However, neither can occupy the runway simultaneously without causing a crash and incurring a heavy penalty.

This two-aircraft landing scenario can be expressed in a normal form game:
\begin{center}
\small
\begin{tabular}{c|cc}
     & $\Occupy$ & $\Yield$ \\ \hline
    $\Occupy$ & $\delta$,$\delta$ & 0,$\rho$ \\ 
    $\Yield$ & $\rho$,0 & $\rho$,$\rho$ \\
\end{tabular}
\end{center}
where $\delta\in\mathbb{R}_+$ is a penalty that is arbitrarily larger than any other values in this matrix and $\rho\in\mathbb{R}_+$ is a small penalty that occurs when a player chooses the $\Yield$ action. 
The cost matrix for this game is as follows:
\begin{equation} \label{eq.ATMCostMatrixGeneral}
    C^{ij}=C^{ji}=
    \begin{bsmallmatrix}
        \delta & 0 \\ \rho & \rho
    \end{bsmallmatrix}
\end{equation}
The best outcome, 0, occurs when each player occupies while the opponent yields. The worst case occurs when both players $\Occupy$, thus resulting in $\delta$ penalty for each player.

This game setting can be expanded to various air traffic management scenarios. Players could be airlines competing over airport slots, departure and arrival queues competing over runway usage, and airports competing over an air route.

\subsection{Airport departure/arrival queue game}
We will focus on the scenario of the departure and arrival queues competing over the limited number of runways, as illustrated in \Cref{fig:introimage}.
\subsubsection{Players}
Let \emph{queues} indicate a queue of aircraft waiting for departure or arrival. A queue that holds aircraft waiting for departure is \emph{departure queue}, and one with aircraft waiting for arrival is an \emph{arrival queue}. An aircraft is added to the $i$-th queue according to a Poisson process at a specific rate $\poisRate_i$ for each queue, $\mathrm{Pois}(\poisRate_i)$. Each departure or arrival queue is a player in this game who wants to reduce their queue length by deploying the aircraft in their queue to occupy runways.

\subsubsection{Actions}
The game is played every 5 minutes. Every time the game is played, the players have two actions for each runway: Occupy the runway or yield its usage to the others. Each player can decide whether to occupy or yield each runway independently if multiple runways exist. Let $r$ be the number of runways; then the number of actions, $m$, becomes $m=2^r$. For example, if there are two runways, then there are four possible joint actions: $\actprofset=\{$($\Occupy$, $\Occupy$), ($\Occupy$, $\Yield$), ($\Yield$, $\Occupy$), ($\Yield$, $\Yield$)$\}$. In practice, the control tower (coordinator) will choose among these joint actions based on $\jointz$ and then radio a recommended action to each aircraft in the front of each queue.

\subsubsection{Cost matrix}

The cost for each player (queue) in this game represents the delay applied to all the aircraft within the queue. Since the game is played every 5 minutes, if a player chooses to $\Yield$, it causes 5-minute delays to all the scheduled aircraft in its queue. We designate 5 as $\rho$, the $\Yield$ penalty. As the average number of added aircraft to each queue per unit time $i$ is proportional to $\poisRate_i$, the cost for $\Yield$ becomes $5\poisRate_i$, or $\rho\poisRate_i$ minutes per runway. If a particular player chooses $\Occupy$ and all other players do $\Yield$ the runway, that player does not incur any costs. However, if the simultaneous occupation of a runway occurs, all the occupying players receive a large penalty $\delta \gg \rho$.

An example cost matrix for queue $i$ when there are 2 runways, or $m=4$ is,
\begin{equation}
    C^{ij}=\poisRate_i
    \begin{bsmallmatrix}
        2\delta & \delta & \delta & 0 \\ 
        \delta+\rho & \delta+\rho & \rho & \rho \\
        \delta+\rho & \rho & \delta+\rho & \rho \\
        2\rho & 2\rho & 2\rho & 2\rho
    \end{bsmallmatrix}
    , \forall j \in -i
\end{equation}
where the order of actions is ($\Occupy$, $\Occupy$), ($\Occupy$, $\Yield$), ($\Yield$, $\Occupy$), ($\Yield$, $\Yield$).

\section{Numerical Experiments}
We evaluate the effectiveness of the RRCE algorithm by comparing it with the algorithm based on the correlated equilibrium and Nash equilibrium, respectively \cite{juliaPATH}. The algorithm based on the correlated equilibrium, termed the \emph{CE algorithm}, solves the optimization in \Cref{eq:CorrOptimization}. The algorithm based on the Nash algorithm termed the \emph{Nash algorithm}, finds a solution that satisfies the conditions in \Cref{eq:NashKKT}. It converges to a different equilibrium under different initialization when multiple Nash equilibria exist.

The proposed RRCE algorithm has two variations depending on which strategy it uses to find multiple Nash equilibria; RRCE using the \emph{random initialization} is denoted as the \emph{Random-RRCE}. It repeats \emph{random initialization} method for a fixed number of times, and the newly discovered Nash equilibrium is returned. The RRCE using the \textit{brute-force} is denoted as the \emph{Brute-RRCE}. It performs an exhaustive search for \emph{pure} Nash equilibrium (with a deterministic strategy) by enumerating the best responses of each player for all joint actions.

In summary, four algorithms are used for the experiment (two RRCE algorithms, the CE algorithm, and the Nash algorithm). Note that other baseline algorithms are neglected since there have not been any algorithms that share common objectives with the proposed algorithm, making the direct comparison unnecessary, per the discussion in \Cref{section:RelatedWorks}.

To capture the variance in computation time and solution quality of each algorithm, we conduct a series of Monte Carlo experiments. Fifty trials with fixed arrival rates $\poisRate_i$ are tested for each distinct number of players ($n$) and actions ($m$). A total of 18 settings for ($n$, $m$) are investigated, with $n$ varying from 2 to 7 and the number of runways $r$ varying from 1 to 3, which is equivalent to the number of actions, $m\in\{2,4,8\}$ respectively. These test cases correspond to the number of joint actions ranging from $2^2$ to $2^{21}$. We implement algorithms in the Julia programming language, using the ParametricMCPs.jl package \cite{juliaParametric} and the PATH mixed complementarity program solver \cite{juliaPATH}. All experiments were performed on an AMD Ryzen 9 7950X processor.

\subsection{Evaluation criteria}
We use three metrics to measure the performance of each algorithm: the computation time, the Gini index, and the average delay cost per player.

\subsubsection{Computation time (CT)}
The computation time measures the time required to compute the solution. We evaluate the computation time in two ways: (i) \textit{Solver time}, which is the runtime for the solver to find the solution, and (ii) \textit{Total computation time}, which is a combination of the solver time and the \textit{pre-processing time} required to compile the problem before running the solver.

\subsubsection{Average delay cost (AC)}
The average delay cost measures the average cost of the players as
\begin{equation}
    \textstyle \mathrm{AC}(\textbf{c}) = \frac{1}{n}\sum_{i\in\IntRange{1}{n}}c_i,
\end{equation}
where $\textbf{c}=\{c_1,c_2,\ldots,c_n\}$.

\subsubsection{Gini index (GI)}
The Gini index is an indicator measuring fairness among the players. The smaller the discrepancy in the cost between the players, the better the fairness. A small Gini index indicates the players achieve a more fair outcome. We compute the Gini index as
\begin{equation}
    \textstyle \mathrm{GI}(\textbf{c}) = \frac{1}{2\mathrm{AC}(\textbf{c})n^2}\sum_{i,j\in\mathbb{N}_{[1,n]}}|c_i-c_j|,
\end{equation}
where $\mathrm{AC}(\textbf{c})$ is an average delay cost computed from \cite{FairnessGuide}.

\subsection{Objective function}
There are two objectives that the coordinator in this scenario has to satisfy: (i) Minimize the total cost and (ii) Minimize the cost differences between the players. Qualitatively speaking, the first objective promotes the overall system's efficiency, and the second objective promotes fairness among the players.

We adopted a cost function that balances both objectives: the \emph{fairness-threshold-criteria with maximin fairness} \cite{FairnessGuide}. It is formulated as,
\begin{equation} \label{eq:fairThresCrit}
    \textstyle J(\jointz) = -n\Delta+\sum_{i=1}^n \max(c_i+\Delta,\textbf{c}_{\max}),
\end{equation}
where $\Delta$ is a pre-specified fairness-threshold value and $\textbf{c}_{\max}$ is the maximum value of \textbf{c}. Note that each $c_i$ can be computed from $\jointz$ directly via \Cref{eq:cost_with_Z}. When the differences of costs between all the players are within $\Delta$, the cost function becomes $J(\textbf{c})=\sum_{i=1}^nc_i$. If a single player incurs a higher cost than any other player by more than $\Delta$, the overall cost depends only upon that player's cost, and $J(\textbf{c})=-n\Delta+n\textbf{c}_{\max}$.
% If the difference between player $i$'s cost and $\textbf{c}_{\max}$ is less than $\Delta$, the cost function adds $c_i$ into consideration. However, if the difference is larger than $\Delta$, or $c_i$ is less than $\textbf{c}_{\max}$ by more than $\Delta$, the cost function focuses on $\textbf{c}_{\max}$ rather than $c_i$. 

\subsection{Result}
The RRCE and Nash algorithm solve test cases involving up to $2^{21}$ joint actions. However, the CE algorithm fails to solve several test cases due to a lack of available memory, thus leaving missing data points in the figure. The maximum number of joint actions the CE algorithm can handle is $2^{9}$.
\begin{figure}
    \centering
    \includegraphics[width=0.48\textwidth]{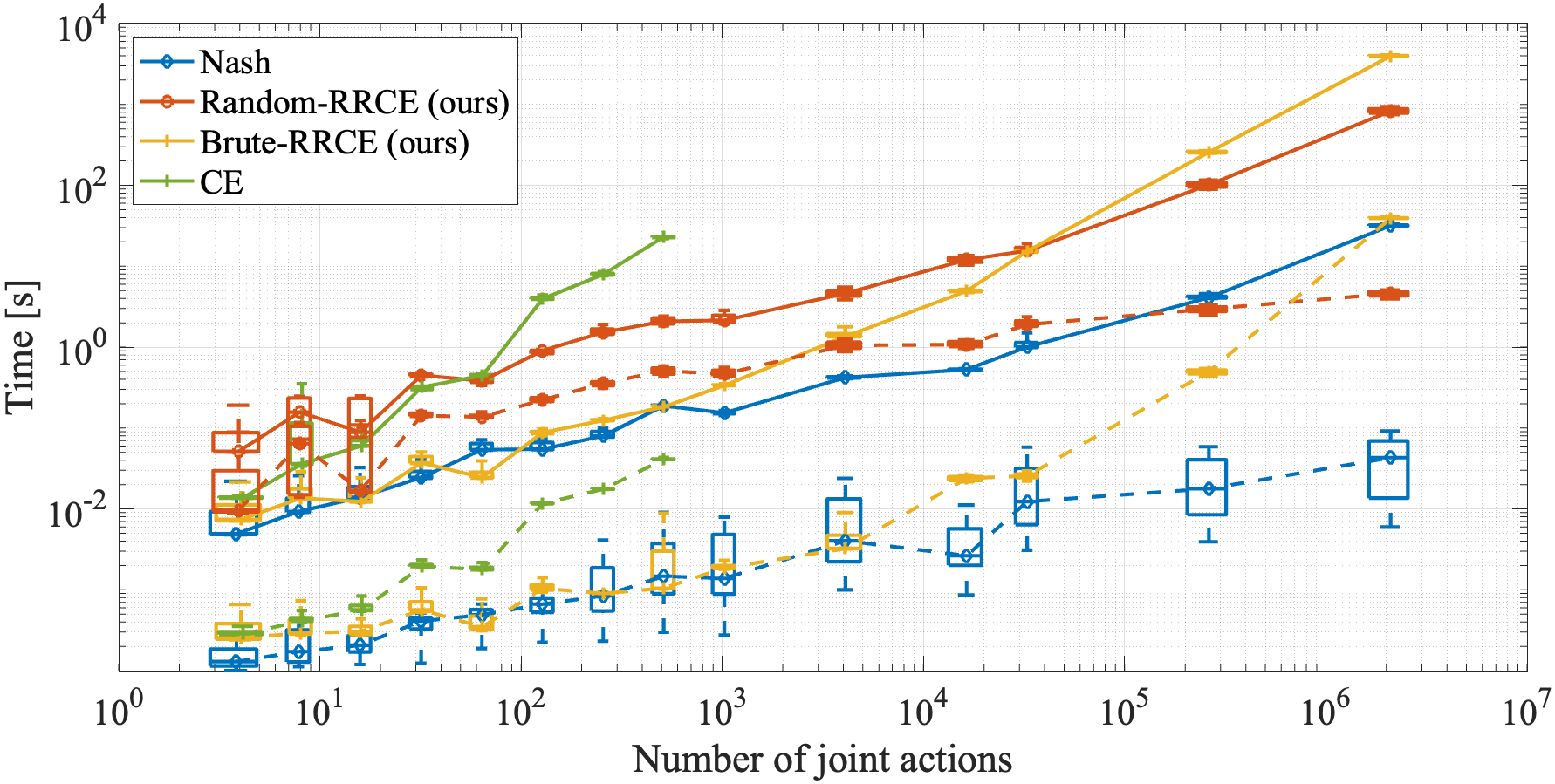}
    \caption{Computation time plot in log-log scale. Solver runtime plot (dotted) and total computation time (solid) that includes both solver runtime and preprocessing time.}
    \label{fig:CT}
\end{figure}
\begin{figure}
    \centering
    \begin{subfigure}[b]{0.48\textwidth}
        \includegraphics[width=\textwidth]{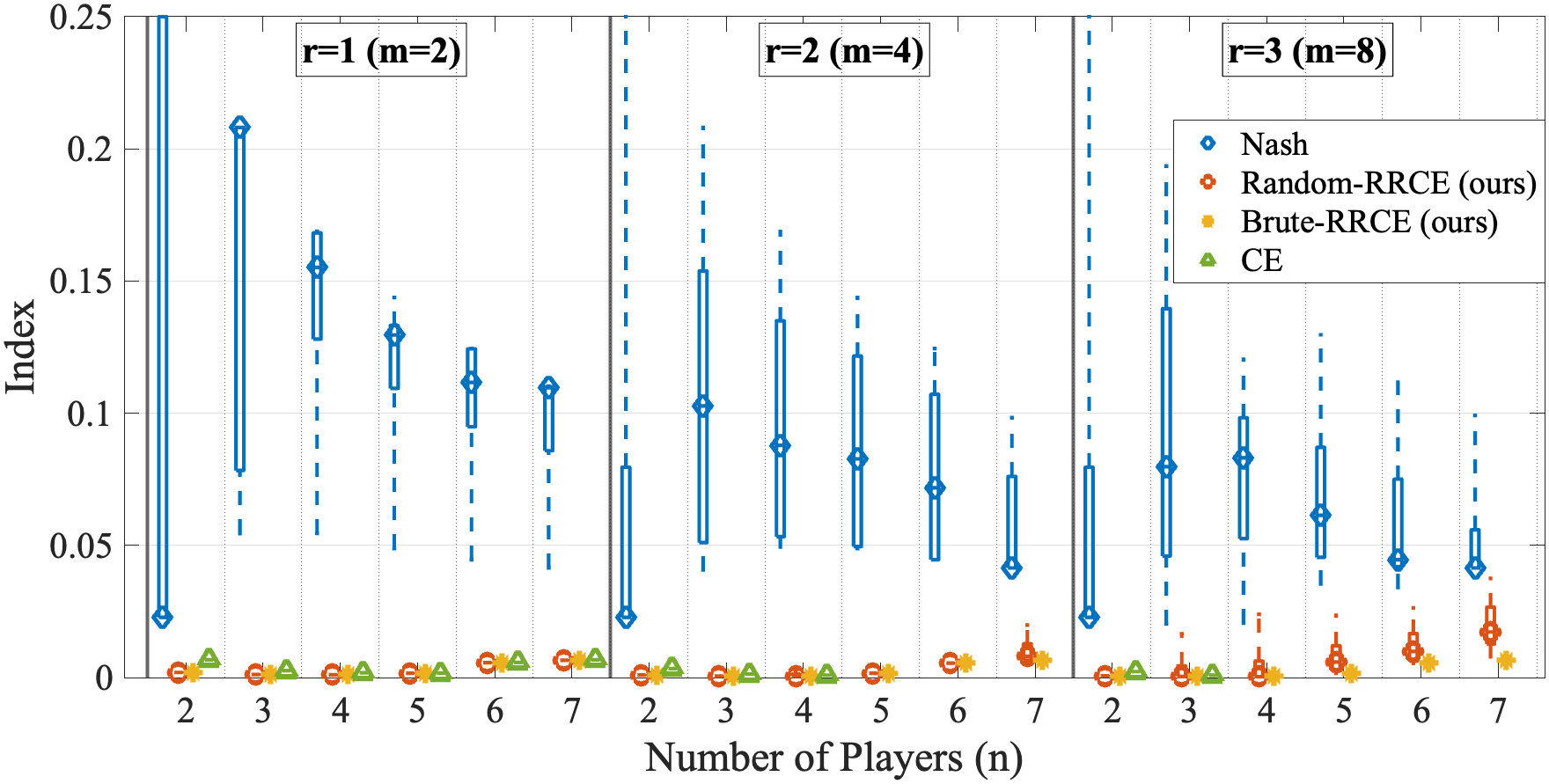}
        \caption{Gini index} \label{fig:gini}
    \end{subfigure}
    \begin{subfigure}[b]{0.48\textwidth}
        \includegraphics[width=\textwidth]{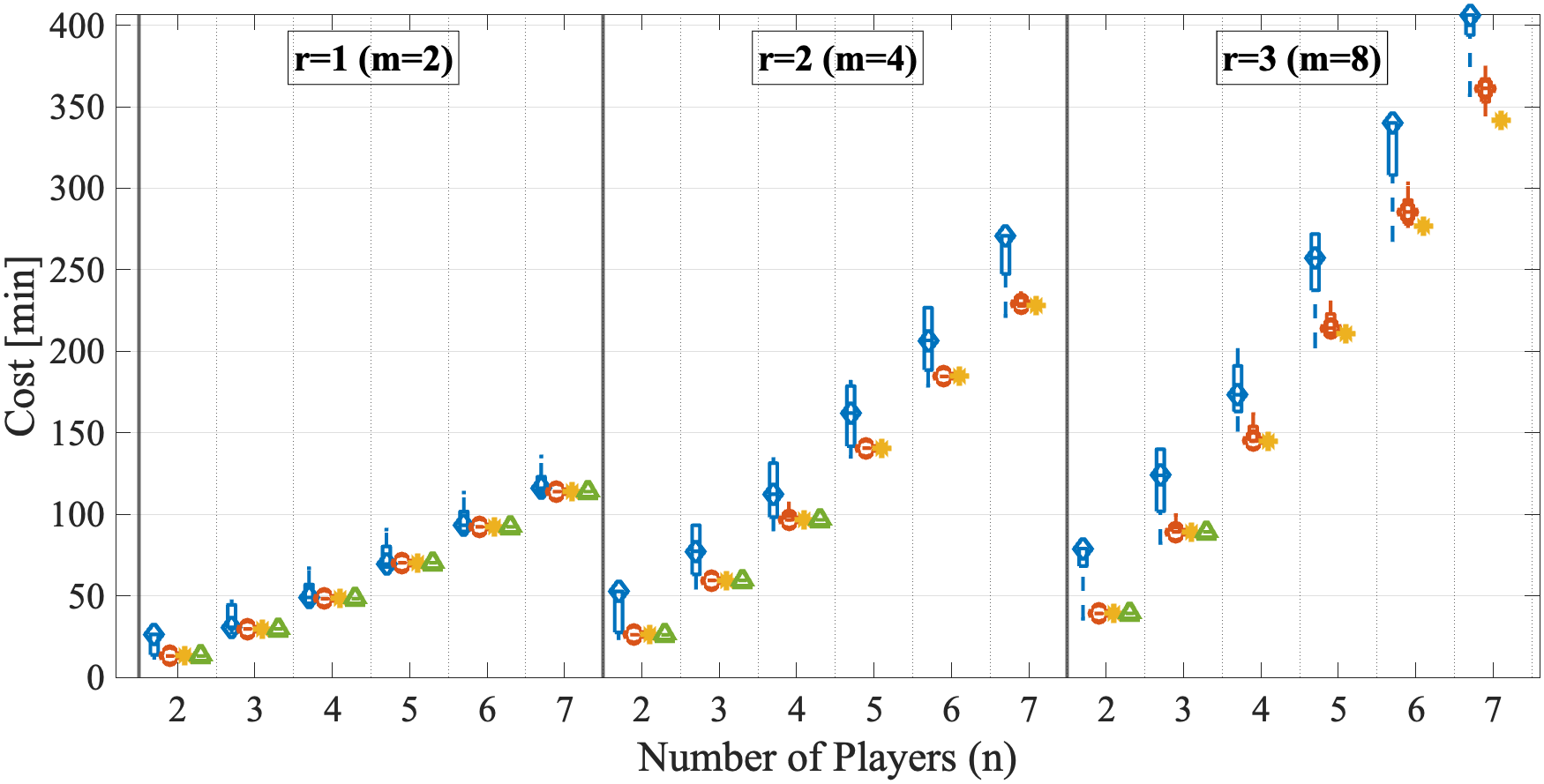}
        \caption{Average delay cost} \label{fig:avgdelay}
    \end{subfigure}
    \caption{Unfairness indicator (Gini index) plot (upper) and average delay cost plot (bottom). Results are shown per runway number $r$ and the number of players $n$.}
    \label{fig:SQ}
\end{figure}
\subsubsection{Computation time}
\Cref{fig:CT} shows the computation time regarding the number of joint actions in each test case. Random-RRCE shows the largest value in terms of solver time for most of the test cases. However, it shows a polynomial increasing trend, which is also true for the Nash algorithm. The Brute-RRCE and CE algorithms show a much greater rate of increase in solver time, and the CE's rate of increase is the greatest (\ie\ worst) among the algorithms.

A similar trend exists in the total computation time. The Random-RRCE and Nash show polynomial increasing trends, while Brute-RRCE and CE show a greater rate. The total computation time for CE exceeds that of the Random-RRCE in the test case involving 64 joint actions, becoming the most time-consuming algorithm. The maximum total computation time reduction observed from Random-RRCE to CE was 91.0\% in a test case involving $2^{9}$ joint actions.

\subsubsection{Gini index}
\Cref{fig:SQ} shows the solution quality indicators (Gini index, average delay cost) with respect to the number of players ($n$) and actions ($m$). Regarding fairness, \Cref{fig:gini} shows that the RRCE and CE algorithms kept the costs between the players similar. Among the algorithms apart from the Nash algorithm, the Random-RRCE shows the highest (worst) median Gini index value and higher variance, and the trend becomes notable in the cases involving larger $n$ and $m$. This degradation is due to the reduced ratio of the number of sampled Nash equilibria to the number of all Nash equilibria in the game as the problem size increases. 

\subsubsection{Average delay cost}
According to \Cref{fig:avgdelay}, the median of average delay costs for CE and RRCE algorithms are consistently lower than that of the Nash equilibrium. RRCE and CE yield a solution showing similar performance to the cases CE solved. Random-RRCE shows performance between these two extremes, with the average delay cost reduced by 1.8\% up to 50.4\% than that of the Nash approach, while showing a maximum optimality gap of 0.066\% compared to the observable CE algorithm results.

\section{Conclusion and Future Works}

We propose a highly scalable algorithm that computes the correlated equilibrium, a coordination mechanism that has been proven helpful in multiplayer noncooperative games, by approximating the set of correlated equilibria with the convex hull of multiple Nash equilibria. Numerical experiments show that the proposed algorithm demonstrates improved scalability compared to the standard correlated equilibrium computation and superior solution quality (fairness and average cost per player) to the Nash solution, which does not involve coordination. 

Despite this promising result, there is room for further development. As this paper focuses on the correlated equilibrium set approximation method, we adopted a simple strategy to search for multiple Nash equilibria. However, as the problem scale increases, the Nash equilibria recovered represent a decreasing fraction of the total number of Nash points, which makes the convex hull approximation worse. Thus, exploring a method that can search for Nash equilibrium to maximize the volume of the convex hull effectively will be an exciting topic for future study.

\bibliographystyle{IEEEtran}
\bibliography{IEEEabrv,reference}

\end{document}